\documentclass[12pt]{article}
\usepackage[dvips]{graphicx}

\begin{document}
\begin{center}
\large {\bf Microscopic Discrete Proliferating Components Cause the Self-Organized Emergence of Macroscopic Adaptive Features in Biological Systems}\\
\end{center}

\begin{center}
{\large \bf Yoram Louzoun{\scriptsize (1)},Sorin Solomon{\scriptsize (2)}, Henri Atlan{\scriptsize (3)},Irun. R. Cohen{\scriptsize (4)}}  

\textit{\scriptsize (1) Interdisciplinary Center for Neural Computation, Hebrew
University, Jerusalem Israel.  yoraml@alice.nc.huji.ac.il}{\scriptsize \par{}}

 \textit{\scriptsize (2)Racah Institute for Physics, Hebrew University, Jerusalem
Israel.}{\scriptsize \par{}}{\scriptsize \par}

\textit{\scriptsize (3) Human Biology Research Center, Hadassah Hebrew University Hospital, Jerusalem Israel.}{\scriptsize \par{}}

\textit{\scriptsize (4) Department of Immunology, The Weizmann Institute of Science.}{\scriptsize \par{}}

\end{center}
\newpage

\begin{abstract}
We study the emergence of the collective spatio-temporal macroscopic properties of the immune system, by representing individually the elementary interactions between its microscopic components (antibodies, antigens, cytokines).
The results of this detailed explicit analysis are compared with the traditional procedure of averaging over individuals and representing the collective dynamics in terms of densities that obey partial differential equations (PDE).
The simulations show even for very simple elementary reactions the spontaneous emergence of localized complex structures.
In turn the effective dynamics of these structures affects the average behavior of the system in a very decisive way: systems which would according to the differential equations approximation die, display in reality a very lively behavior.
Our conclusions are supported both by explicit microscopic simulations and by analytic calculations.
As the optimal method we propose a mixture of microscopic simulation (MS) systems describing each reaction separately, and average methods describing the average behavior of the agents.
\end{abstract}

\section{Introduction}

The immune system has become lately a classical example of emergence \cite{irun}. 
It is an archetype for biological systems whose behavior cannot be directly deduced from their molecular and cellular properties \cite{complexity}.

As in many other biological systems, a large amount of microscopic data has been gathered. 
This data by itself is insufficient to explain the macroscopic behavior of the immune system.

Like any complex system, the immune one, requires classically a few complementary ways for describing it.
\begin{itemize}
\item The micro-biological level - molecules and cells.\cite{molbio}.
\item The generic properties that give the broad picture of what makes this system tick. For instance the "clonal selection": the random genetic changes of the antibodies and their selection by the affinity with appropriate antigens.\cite{clonal}

\item The macroscopic mechanistic level. For example the aggregated action of groups of cell and cytokine: TH1 vs TH2, idiotypic vs anti idiotypic.... \cite{molbio}

\item The  complex network of interactions. This represents the  emergence of the immune response in terms of a highly connected (e.g neural) network. In particular, one uses neural networks to implement this view \cite{atlan_net,irun_net}.

\item The reaction-diffusion (partial) differential equations level that describes the IS as a nonlinear chemical system \cite{perelson}.
\end{itemize}

We are proposing here a methodology, which has the capacity to express much of the successes of the other methods while transcending their limitations and in particular the limitations which prevented the appropriate understanding of the central issue: the emergence of complex deterministic macroscopic functions out of simple stochastic microscopic interactions .

The methodology is based on recognizing the discreetness and spatial inhomogeneity of the biological units \cite{seiden,simmune} and tailoring a specific model for each biological situation by representing the objects and interactions appropriate for each scale. 

More precisely, we introduce microscopic simulation (MS) \cite{sorin,eldad1,pnas} methods, which on one hand represent directly the basic elements of the immune system but on the other hand allow the identification of the effective macroscopic objects relevant to the collective dynamics of the system.

In our MS we enact in the computer each reaction separately. More precisely we compute at each lattice point the probability of each reaction and then perform reactions randomly according to these probabilities.

We first show that very simple microscopic interactions may lead systematically (when studied with MS) to highly nontrivial effects which one would not have guessed using any of the classical methods.

We describe the resulting new phenomena and show how they are relevant for various biological applications (e.g. the generation of complex spatio-temporal localized self-organized adaptive structures).

In other words we show that systems that would be judged too simple to constitute an adaptive network and too linear to present non-homogeneous solutions in the differential equations context, do in fact emerge well defined/individualized macroscopic objects, which can move around, act in complex ways, and display adaptive behavior.

We present here our Microscopic Simulation and hybrid (discrete-continuum) models as a generic technology for representing in a direct intuitive way the microscopic knowledge on the immune system and for extracting non-trivial macroscopic information out of them (including the emergence of adaptive behavior and self-organization).

Our results are at variance with the predictions of the classical modeling methods \cite{ODE}, but are in encouraging agreement with many known IS features which were long waiting for an explanation.

In chap 2 we study the macroscopic effect of the proliferation of cells induced by a macroscopically homogeneous, but microscopically inhomogeneous signal distribution. We show that the proliferation rate predicted by the MS is many orders of magnitude larger than the one expected from ODE. In particular one obtains proliferation in conditions in which the naive continuum limit predicts decay.

In chap 3 we identify the source of the difference between MS and ODE and propose hybrid models (HM) that bridge the gap between the 2 limits. 

In chap 4 we introduce additional reactions, which limit the growth of the proliferating entities. We show that in certain conditions the macroscopic inhomogeneity is even increased by the effect of competition.

Most of the effects analyzed in the four first chapters are due to the combination of auto-catalysis and of fluctuations. 

In chap 5 We study the effects of discreetization in the presence of negative feedback loops.

We find that some effects predicted by the continuum simulations \cite{osc1,osc2} (but not so frequent in nature) are not present when the system is analyzed using MS (e.g. the spurious oscillatory behavior of certain regulatory pairs).

In conclusion we are opening the way to a wide front re-evaluation of the dynamics that is supposed to emerge out of the known microscopic elementary interactions of the immune system. We provide the main effects to be re-scrutinized and the particular methods to do so.
Paradoxically, the new methods that we propose are closer to the intuition of the practitioners in the immunological field. Indeed the objects which MS manipulates are the very same they encounter in their experimental work: cells, cytokines, epitopes (rather than abstract approximations such as differential equations, neural networks etc).

\section{Dynamics of Discrete Proliferative Agents}
Proliferating entities can be described as agents $A$ that duplicate whenever they meets a stimulating signal $S$. These agents have in general also a typical death rate $d$. 
Examples of such agents are : in immunology - immune cells reacting to an antigen \cite{prol1} ; In ecology - animal proliferating whenever they find food...
 
We will deal with the dynamics of the stimulating signal in the next sections. In this section we will assume the stimulating agent $S$ is fixed in time.

This system can be described as a 2 reactions system (Figure 1).
The first reaction is that whenever am $A$ agent meets a stimulating signal $S$ it multiplies with a probability $d_1$. 
We will note it as :
\begin{itemize}
\item $A + S \rightarrow A + A + S $. $d_1$
\end{itemize}
The second mechanism is the death of an agent with a probability $d_2$
\begin{itemize}
\item $A \rightarrow \Phi$. $d_2$
\end{itemize}

In the homogeneous limit the growth rate of A is proportional to number of A and S pairs. Thus the number of new $A$ agents produced in a time interval $\Delta t$ is:
\begin{equation}
\Delta A = SA*d_1 \Delta T
\end{equation}
The number of A deaths is proportional to the total A population:
\begin{equation}
\Delta A = -d_2*A \Delta t
\end{equation}
Thus the total change in A is:
\begin{equation}
\Delta A  = (S*d_1-d_2)*A \Delta t
\end{equation}
This equation becomes an ODE \cite{calculus} when we take the limit $\Delta t \rightarrow 0$
\begin{equation}
\label{e4}
\dot{A}  = (S*d_1-d_2)*A
\end{equation}
The solution of the ODE $\dot{x}=\alpha x$ is $x=e^{\alpha t}$ \cite{calculus}. In this case the solution is :
\begin{equation}
A = e^{(S*d_1-d_2)*t}
\end{equation}

This solution has 3 regions of behavior (Figure 2):

a) If the average growth rate is higher than the death rate i.e. $(S*d_1-d_2) > 0$ then the $A$ concentration grows with an exponential growth rate of $(S*d_1-d_2)$ This exponential growth will end when other mechanisms, discussed in the next section, become important. 

b) If the average growth rate is lower than the decay rate i.e. $(<S>*d_1-d_2) < 0$ the $A$ concentration decays to 0. 

c) If the average growth rate is equal to the decay rate i.e. $(<S>*d_1-d_2) = 0$ the $A$ concentration is constant in time.

This solution does not take into account the non-homogeneous distribution of $S$. In a real system the proliferation rate changes from one point to the other. Regions with a high $S$ concentration have a high proliferation rate, while regions with a low $S$ concentration have a low (or negative) proliferation rate. 
 
In the ODE we assumed that the effective proliferation rate is equal to the average proliferation rate $(<S>*d_1-d_2)$. This assumption can be very mistaken in many cases. Indeed if the interaction between $S$ and $A$ are local then the local dynamics is described by the local population $A_i$ at each location i.
\begin{equation}\label{e6}
\dot{A_i} = (S_i*d_1-d_2)*A_i
\end{equation} 
The correct averaging of equation \ref{e6} is:
\begin{equation}\label{e7}
\dot{A} = d_1<S_iA_i>-d_2A
\end{equation}

Eq. \ref{e7} transforms into Eq. \ref{e4} if we can separate of $<S_iA_i>$ into $<S_i><A_i>$. Thus in order for Eq. 4 to hold S and A have to be independent variables \cite{basstats}. Eq. 4 is the very basis of any ODE approach. Yet it is obviously wrong in the case we present here since $A_i$ is high precisely in the sites where  $S_i$ is high. Thus S and A are in fact strongly correlated. 

Following this argument we expect to see differences between the microscopic representation of this system and its description by ODE.

Indeed if the signaling agent ($S$) is a random discrete agent its distribution is uniform over large scales but locally its density will contain some granularity. The difference between the large scale $S$ uniformity and its microscopic granularity is the origin of the macroscopic differences between the ODE and the MS results. 

For example if $(<S>*d_1-d_2) < 0$ (the average birth rate is lower than the average death rate) the ODE would yield a decaying solution but the MS yields an exponential growth (Figure 3). 
To understand this growth note that each point in space contains a different concentration of signaling agent ($S$). For most of the points $i$ one will have $(S_i*d_1-d_2) < 0$ and the local population will decay to zero. However for some points  one may have $(S_i*d_1-d_2) > 0$ and the local population at these points will rise with an exponential rate. After a short time the contribution of the points with decaying population will be negligible and only the lattice points with an exponential growth contribute to the average population. In particular the population growth is dominated by the site with the highest $S$ number. 

The previous description does not take into account the diffusion of the $A$'s. The $A$ diffusion will lead to the creation of large scale islands of high A population around each maxima of The $S$ density. These islands will keep growing until they fill all space. The diffusion of the $A$ does not change qualitatively the average population growth mentioned above for a very wide range of parameters \cite{pnas}.
In order to avoid complex formulae we did not represent explicitly the diffusion term $\mu \Delta A$. In all the simulations and analytic computations this term has been dully taken into account. We mention it in the sequel only when it has a major qualitative effect on the results.   

\section{How Well do Different Methods Deal with Discreetness}

The MS helped us to uncover 2 basic features:
\begin{itemize}
\item The population increase in the MS is much higher than the one computed in the homogeneous case: The ODE is exponentially decaying while the MS is exponentially growing.
\item Diffusion by itself is not enough to smooth $A$ concentration distribution.
\end{itemize}

The effect of the non-homogeneity can be observed in every experiment with a petri dish in which the proliferation of cells creates first local patterns of very high concentration even if the original distribution is homogeneous. The irrelevance of diffusion can be observed in many ecological systems, in which although the diffusion rate of each animal is much higher than its reproduction rate localization can still be observed.
 
\subsection{Microscopic Simulation vs. PDE}
One might think that using partial (rather than ordinary) differential equations will appropriately take into account the non-homogeneity in the a distribution.
this is actually not the case: the fact that $A$ can diffuse will only increase the uniformity in the $A$ distribution, while the microscopic inhomogeneity of $S$ is lost upon expressing the $S$ distribution as a continuous function (recall that $S$ was spread with uniform probability across the space).
Consequently there is no reason to expect that the partial differential equations acting on smooth distributions will do better than ODE. This is confirmed by rigorous analytical treatment.

In the next chapter we show that allowing $S$ to diffuse does not solve this problem either:
the diffusion will only interchange one microscopically inhomogeneous configuration with another, equally inhomogeneous one. Their difference as well as their very inhomogeneity is lost once one expresses the distribution of discrete $S$ in terms of a continuous density function.

\subsection{Discreetization Implies Fluctuations} 

The crucial role of microscopic  discreetization in the emergence of the macroscopic inhomogeneity might be unfamiliar and even strange for the reader. After all, the microscopic corrections to the differential equations should morally lead to microscopic corrections in their solutions. That this is however not the case is known in the theory of electric conductivity since the late fifties. Indeed, by solving the Schroedinger equations for periodic potentials one obtains electronic wave functions that spread over the entire space (which implies electric conductivity) \cite{quantum}. However, at least in 2 dimensions, the slightest random uniform noise leads to spatially localized wave functions (insulator) \cite{anderson}. Similarly here the presence of microscopic discreetization of $S$ induces intrinsically microscopic spatial inhomogeneity in the $S$ distribution. The partial differential equation (PDE) for $A$ share the property that (at least in 2 dimensions) the slightest spatial inhomogeneity (of $S$) in the equations leads to macroscopic localization. We have checked that indeed the effect of $S$ discreetness can be fully taken into account by working with un-quantized (real valued) $S$'s with same distribution function as the quantized $S$'s (Figure 4).
continuous
\subsection{Hybrid Models}

The previous section showed that the difference between the MS and ODE results are only due to the microscopic non homogeneity of the $S$ agents. 
We conceived a class of models containing elements from both ODE and MS. We call such models with the collective name of hybrid models (HM). The particular combination of ODE and MS techniques has to be tailored specifically for each biological situation.  

In the present case we compute the local $A$ concentration changes in a continuous way, but use the $S$ distribution produced by the MS. However the spatial distribution of the $A$S is non-uniform, and in fact we don not assume that the $A$'s spatial distribution is .

In other words we replace the stochastic dynamics of the MS by one differential equation per lattice site, and we preserve thereby the inhomogeneity in the agents concentrations. 

Other system require other HM that fit their specific aspects. For example a system containing regions with high interaction rates, and high concentrations, and regions with low interaction rates can be modeled using a HM that computes the interaction rate stochastically for low concentrations, and deterministically for high concentrations.

HM model has the advantage that when the number of A and S pairs per site is large, the local reaction rate can be computed directly rather than having to enact explicitly each and every $A+S->A+A+S$ reaction.  Thus they have a much lower CPU cost than the MS.

intrinsically This very simple system shows that the results one obtains depends crucially on the appropriate use of models, and the recognition of the limits of applicability of the various approximations.
It draws in rough lines the limits between the range of applicability of each type of model. 
\begin{itemize}
\item ODE can be used when the diffusion rate is much higher than the maximum activation rate.
\item MS are precise over a very wide range but have a very high cost in CPU time. 
\item 4D PDEs have all the severe problems related to the continuity assumption, and are not much cheaper than MS. However there is a small range of cases where they can provide analytical solutions \cite{PDEsol}.
\item HM can reproduce most of the results produced in MS without the high CPU cost. They reproduce the effects of microscopic inhomogeneity. 
\end{itemize}

The HM direction can be used in an adaptive way. In particular the collective objects emerging  form the dynamics can be treated as the effective elementary agents for higher (coarser scale) level of analysis.

\section{Mechanisms Limiting the Population Growth}
The proliferation of the $A$ agents in the models of section 2 must be limited by some mechanism if we do not want them to take over the universe. 
We have shown that a high death rate is not enough in order to limit the population proliferation. 
The required mechanisms can be either an external regulator \cite{regul}, or a limit on the resources available to the proliferating cells \cite{limit}. 

\subsection{Local Competition}
Many of the mechanisms limiting the A proliferation can be described as competition. 
Competition for a resource means that whenever 2 cells need the resource either to survive or to proliferate only one of them will get it and the other will die (or fail to proliferate).
In ecology competition can be over food or water, while in the immune system competition can be over access to an antigen (Figure 5).

Such a system is sometimes called a Lotka Voltera system \cite{LV1,LV2} and can be described using a local differential equation:
\begin{equation}
\label{loc}
\dot{A}=d_1S_i*A_i-d_2A_i-\lambda A_i^2.
\end{equation} 
The system can exist in two regimes:
\begin{itemize}
\item If the proliferation rate is higher than the death rate, this system will proliferate until it reach the saturation level uniformly throughout the entire space. In such a case there is no difference between the MS and ODE results.
\item If the proliferation rate is lower than the death rate than the ODE would eventually converge to a uniform vanishing concentration. By contrast the MS will reach a steady state  of islands whose population density is at saturation level. These islands are embedded in a background empty of $A$. For high $A$ diffusion rate the MS yields non-zero concentration everywhere (Figure 6) in spite of the ODE prediction. 
\end{itemize}

Paradoxically the conditions in which  ODE and the MS show similar features are seldom realized in nature. They represent extreme cases in which a certain agent is capable to fill all the available space. In the animal world only a few species enjoy such a status (For example humans and their parasites). In immunology this regime corresponds to a cell type filling all the blood and immune system (systemic diseases). 
Except for these extreme cases the situation in which  $(<S>*d_1-d_2) < 0$ seems to be more appropriate for realistic systems.

\subsection{Global Competition}
The local competition mechanism presented in the previous section assumes that the limiting factor that creates the competition is homogeneous on the same scale as the $A$ agents concentration. (The scale of homogeneity is the spatial scale in which a cell type has a homogeneous concentration). 
Some reactions leading to competition have a larger homogeneity scale than the factor they are limiting. For example if cells proliferate and require for their proliferation a substance found in the blood.
In other words, while the proliferation is a local mechanism, the factor leading to competition is a global one . 
In ecological systems the global character of the limiting factor may be related with the fact that the predators move much faster than their pray and that their population is proportional to the total prey population. Therefore their population constitutes an indirect interaction between distant prey locations. We will show that in such cases the effects of inhomogeneity are even more important than in the previous sections. As an aside let us note that this kind of reactions although formally local (A lion can eat a pray only if they meet) are very effective in inducing top down effects; that is the regulation of the local pray population according is enforced by their total population. 

The overall homogeneity and uniformity assumption used in ODE has the benefit of merging into one formalism a large number of biologically different homogeneous systems. This advantage turns into a disadvantage when multiple spatial scales are involved in the system, as will be described in this section. 
MS on the other hand can describe in a natural way only local interactions. It is non-trivial to describe with MS factors acting on widely different scales. The solution in this case would be to create a new type of HM, which will allow in particular to express the interaction between agents and their averages over various ranges.

A simple example for such system is obtained by modifying the regulation mechanism of the system presented in the previous section (Figure 7). There the 2 mechanisms proliferation and competition were local interactions. We can now replace the local competition by a global competition. Even-though from the microscopic point of view the non-locality might look unnatural this is the only way that nonlinearities can arise at the macroscopic level. Indeed A term of the form $<A>^2$ can appear only as the average over terms of the type $A_i*<A>$. Therefore the modification of Eq. \ref{e6} would be  
\begin{equation}
\label{lv}
\dot{A_i}=d_1 S_i A_i- d_2 A_i- \lambda A_i <A_i>
\end{equation}
That is we replaced the local competition in the system described in section 4.1 by a global competition term. Once again we find that passing from local to global reactions has crucial implications: By comparing the behavior of the system Eq. \ref{lv} with the system Eq. \ref{loc} one find that they are dramatically different in all the parameter regions.
Even in the region in which the system Eq. \ref{loc} is described correctly by an ODE, the ODE are still incapable to describe properly the system Eq. \ref{lv}. Indeed in the \ref{lv}  system the entire space is empty of $A$s except for a single island centered around the point in which the $S$ concentration is maximal (Figure 8). 

We can understand this system by simplifying it to :\footnote {We rescaled the equation to $d_1=1, \lambda=1$ and added the death term $d_2$ into the $S$ density. This can be done with no loss of generality}
\begin{equation}
\label{comp}
\dot{A_i}=A_iS_i-A_i<A_I>=A_i(S_i-<A_i>)
\end{equation}

One can see that in every point in which $S_i$ is lower than $S_{max}$ the $A_i$ density will vanish, while around the point $i_{max}$ in which $S_i=S_{max}$ a region of high $A_i$ density will appear. 
Suppose that one starts with all the $A_i$s very small; 
As long as the average concentration of the $A$'s ($<A_i>$) is lower than $S_{max}$, the local concentration $A_i$ at the point $i_{max}$ will rise (acording to Eq. \ref{comp}). This in turn will raise the value of $<A_i>$.
Eventually $<A_i>$ will reach the value of $S_{max}$. What will be the equilibrium value of the $A_i$s in every other point?
In every point where $S_i < S_{max}$, one will have $A_i(S_i-<A_i>)=A_i(S_i-S_{max})<0$, which according to \ref{comp} means that the concentration at those points will decay to 0.
It is clear therefore that the entire $A$ population responsible for $<A_i>=S_{max}$ is due to a very dense island situated around $i_{max}$. 
The equality  $<A_i>=S_{max}$ is not affected by the  diffusion, which only has the effect of spreading the distribution of the $A_i$'s in a limited region around $i_{max}$. 
One sees that the dynamics is dominated in this case by the most extreme stochastic local fluctuation in the $S_i$ distribution, and not by the global or average properties of the $S_i$s.

\subsection{Dynamical $S$s}
In the previous section we considered a static $S_i$ distribution.
The intrinsic microscopic inhomogeneity of the $S$s produced large-scale inhomogeneity in the concentration of the $A$s. 
We will show that these effects survive the introduction of $S$ diffusion, as long as the maximal proliferation rate of the $A$'s is not lower than the diffusion rate of the $S$s. 
If the signaling agent diffusion rate of the $S$ is very high, the $A$s will ``see'' a blurred $S$ concentration and will react only to the average $S$ concentration, as in the ODE limit. 
If however the diffusion rate of the $S$ agents is not much higher than the $A$s proliferation rate then the $A$s do react to the local concentration of the $S$s. 
The $A$s produce peaks in their concentration at the current locations of the $S$ maxima. The locations of the $S$ maxima change since the $S$s diffuse. Consequently the regions of high $A$ concentration will follow them. This will lead to long term large (intermittent) fluctuations in the total $A$ concentration (Figure 9.).

In this system the difference between the ODE and MS is not due to any imposed inhomogeneity in the starting conditions. It is only due to the discreteness of the $S$s.

\subsection{Emergence of Complexity}
In the first chapters it has been shown that the discrete character of the elementary agents $S$ leads to unexpectedly rich and complex behavior of the $A$s population. The complex spatio-temporal emerging structure of the $A$ population cannot be predicted by an ODE approach. Moreover in this approach one cannot even ask the appropriate questions about the system's dynamics. However a system which has only proliferation and death (linear) terms leads eventually to a divergent $A$ population. 
  In the present chapter we introduced a few types of competition terms. Local competition limits the size and maximum concentration of each macroscopic $A$ island. However rather than interfering with the emergence of macroscopic features, the competition terms turned out to enhance the local and complex character of the $A$ islands.
Global competition can endow these macroscopic islands with an emerging social (or sometime anti-social) behavior. The evolving macroscopic objects may now have apparent ``goals'': seeking food, multiplying, dying or destroying the competing islands. 
The lifespan of these emerging islands is many order of magnitudes longer than the lifespan of their components (which of course remain non adaptive and totally mechanical during the entire history of the system). 

\section{Negative Feedback Loops}
\subsection{The Survival of the Weakest}
The complex behavior observed in the preceding sections is the result of positive feedback loops (autocatalyis). One could think that the difference between the ODE and MS descriptions is specific to systems with autocatalytic components. We will show in this section that this is not the case.
In fact we consider a simple negative feedback situation: The response of the immune system to a pathogen invasion. 
The high pathogen concentration leads to an increase in the immune cells concentration which in turn leads to a decrease in the pathogen concentration.

One can express the dynamics of this system using ODEs, in which $x$ is the concentration of the pathogen and $y$ is the concentration of the immune cells.
\begin{eqnarray}
\label{path1}
\dot{x}=-axy\\
\label{path2}
\dot{y}=bxy-cy=(bx-c)y
\end{eqnarray}
Eq. \ref{path1} expresses the fact that upon meeting an immune cell the pathogen is destroyed. Eq. \ref{path2} expresses the fact that the immune cells proliferate when they meet a pathogen with a probability ($b$), and die with a constant rate ($c$).
The solution of the ODE \ref{path1},\ref{path2} consist in a fast rise of the immune cell population, and following it a decrease in the pathogen population, until the pathogen population is lower than ${c \over Oscillationsb}$. After that point the immune cells population starts to decrease. The dynamics of the MS and the ODE are the same in this case, with a very notable difference in the final steady state of the system:
\begin{itemize}
\item Using ODE one gets 2 types of asymptotic steady states depending on the initial condition ($x(0),y(0)$). More precisely if  $bc*(ln(bc)-1)-ay(0)+bx(0)-c*ln(bx(0))<0$ both the immune response and the pathogen decreases asymptotically to zero. Otherwise the immune response asymptotically decreases to zero, while the pathogen population approaches a non-zero asymptotic value. This can be interpreted as a constant latent infection. 

\item In the MS various spatial regions may converge either of the 2 steady states described above within a finite time.  The regions containing a non zero concentration of pathogens (and no immune cells) constitute protected reservoirs of pathogens.

This can explain for example the fact that even people receiving potent anti HIV drugs for which the average concentration of HIV is undetectable for a long period reproduces the disease spontaneously when the drugs are stopped. 
\end{itemize}

\subsection{MS Eliminates Spurious Oscillations}
It has been shown in the previous sections that ODE fail to represent emergent features of dynamical systems, but ODEs can also generate artifacts that do not appear in real systems. 
Such artifacts are exemplified by the following system consisting 3 cell types: An external antigen, idiotypic (Id) and suppressor cells (A-Id).
The antigens stimulate the Idiotypic cells. The idiotypic cells in turn stimulate the anti idiotypic cells, which destroy idiotypic cells. We also introduce a saturation term that regulates the anti idiotypic cell population (Figure 12). 
The ODE describing this system is :
\begin{eqnarray}
\dot{x}=ax-bxy\\
\dot{y}=cxy-dy
\end{eqnarray}
where x is the concentration of the Id cells and $y$ is the concentration of the a-Id cells. Such equations have an oscillatory solution around $x={d \over c}, y={a \over b}$ with a period of $\sqrt{da}$ (Fgigure 13).

The same reactions simulated at the microscopic level do not produce oscillations \cite{lotkanadav}. The MS produces regions containing a high concentration of Id cell with very low concentration of A-Id cells around them, or regions of A-Id cells that completely destroyed the Id cells around them dominate the concentration. The disappearance of the oscillations in the MS is due to the de-phasing of the different regions in space. Each regions variate independently so that the average yields a constant concentration. When one cluster has a maximum concentration of Id cells, other clusters can have a low concentration of Id cells.
Therefore the oscillatory character of the ODE solution is related with the unrealistic assumption that the system is precisely homogeneous in space.

If the concentration of the Id and Anti Id are low, one can still have in the MS system quite significant fluctuations around the average due to the finite size of the system relative to the clusters size (Figure 14) (i.e the fact that the system contain at every moment a small number of clusters, so that the variation of every cluster influences the average behavior). This fluctuations involve the total disappearance of agents in certain regions of space. In such a situation it might take a macroscopic amount of time until this region is ``colonized'' again by agents immigrating from other clusters. 

This simple system shows that the appearance of oscillations cannot occur spontaneously, unless there is an explicit mechanism phasing the whole system. 
This mechanism can be a very high diffusion and/or an external input forcing a unified timing over the whole system.Oscillations appear in many descriptions of systems using ODE \cite{osc1,osc2}. This simple system shows that these oscillations are many times only an artifact of the model continuity assumption, and not a realistic feature of the system.

\section{Discussion}
Dynamic Biological systems were traditionally described using ODE. 
ODE were invented and proved useful in simple physical systems that fullfiled certain restrictions:
\begin{itemize}
\item A very large density of interacting particles at every point in space. 
\item A high reaction rate.
\item A diffusion rate higher than the local interaction rate.
\item An immediate result of each interaction. i.e. no delay between the interaction and its result.
\end{itemize}
All of these assumptions are usually false in biological systems.
\begin{itemize}
\item The density of interacting particles is small. For example the number of interacting cells at every point is of order of tens. 
\item The reaction rate is low - The interaction between cells is a long process involving a long chain of molecular sub processes.
\item  The diffusion rate can in many cases be low. The motion of cells cannot be described as a diffusive process. The motion is best described as a slow directional motion directed by many chemical signals.
\item Most biological processes require the production of new molecules or the division of cells. Each of these process can take hours - days. Thus there is a delay between an action and its result.
\end{itemize}
Moreover the dynamics of physical systems is dominated and in fact determined by conservation laws. In biology none of the classical quantities (mass,number,energy,momentum....) is conserved.

In this work we showed that even in simple cases ODEs fail to describe appropriately the dynamics. This is as well: Systems which were too simple to present complex emergent features in the framework of ODE turned out to present when studied appropriately a wide range of experimentally documented features .
We used in addition to the straight-forward MS intermediate techniques (HM) that allow to take explicitly into account the intermediate relevant dynamical scales between the MS and ODEs.

\subsection{Modeling Autocatalysis}
One of the main sources of complexity is the combination of microscopic inhomogeneity and autocatalysis \cite{chaos1}. In biological systems there is an inherent source of microscopic inhomogeneity- the discreetness of the entities composing biological systems.
 Moreover most biological systems involve proliferation which is the basic autocatalitic mechanism. In reality most situations are mixed containing both autocatalytic and regulatory mechanisms and can be modeled using custom made models describing globally the non catalytic parts and explicitly the microscopic details of the autocatlytic parts.

\subsection{Inter-Scale Information Flow}
Autocatalysis results in the flow of information from small scale to large-scale feature \cite{autocatalsys}; from the microscopic details of the system to its macroscopic behavior. The rapid evolution of small-scale features to a macroscopic scale affects the distribution of the macroscopic variables. In particular the macroscopic variables which are dominated by the sum of a small number of large values (instead of on the sum of a large number of small values) have a very specific (fractal-power law) space-time behavior. Such effects are enhanced when there is a direct effect of the macroscopic values on the microscopic mechanisms. Such a dialog between different scales is the obvious reality in biological systems \cite{scaledialog}. 

Many biological systems evolve at time scales much longer than their fundamental microscopic time scales. In ODE the emergence of long time scales cannot occur naturally, while in MS they can emerge spontaneously.
In MS, the agents self-organize \cite{atlan} in collective structures with emergent adaptive properties and with dynamical space-time scales of their own.
The effective dynamics of these emerging objects is much slower than the one of their components (and this fact constitutes their very definition \cite{critica}). Analyzing the system at the level of the emerging objects require automatically the transition to longer time scales and larger spatial scale. Each level corresponds to a new space-time scale in a hierarchy of complementary representations of the same complex macroscopic system. 

We found in this paper that the core mechanism for the emergence of collective macroscopic objects is the interplay between autocatalysis and the discrete microscopic structure of the components of the system. This mechanism is totally missed by the mean field description, which predicts uniform extinction in conditions in which the real system emerges actually very lively collective adaptive objects.

\newpage

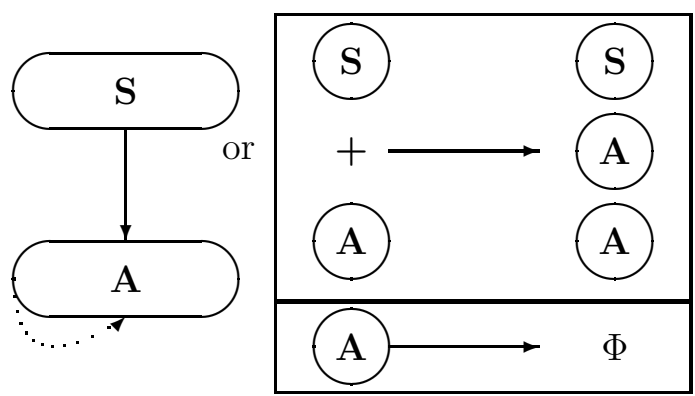
\begin{figure}
\setlength{\unitlength}{1cm}
\begin{picture}(10.5,5.0)

\thicklines

\put(3.0,4.0){\oval(3.0,1.0)\makebox(0,0){\bf \large {S}}}
\put(3.0,3.5){\vector(0,-1){1.5}}
\put(3.0,1.5){\oval(3.0,1.0)\makebox(0,0){\bf \large {A}}}
\bezier{12}(1.5,1.5)(1.5,0.0)(3.0,1.0)
\put(3.0,1.0){\vector(1,1){0}}

\put(4.5,3.2){\makebox(0,0){\large {or}}}

\put(5.0,1.2){\framebox(5.5,3.8){}}
\put(6.0,4.4){\oval(1.0,1.0)\makebox(0,0){\bf \large {S}}}
\put(6.0,2.0){\oval(1.0,1.0)\makebox(0,0){\bf \large {A}}}
\put(6.0,3.2){\makebox(0,0){\bf \large {+}}}
\put(9.5,4.4){\oval(1.0,1.0)\makebox(0,0){\bf \large {S}}}
\put(9.5,3.2){\oval(1.0,1.0)\makebox(0,0){\bf \large {A}}}
\put(9.5,2.0){\oval(1.0,1.0)\makebox(0,0){\bf \large {A}}}
\put(6.5,3.2){\vector(1,0){2}}
\put(5.0,0.0){\framebox(5.5,1.2){}}
\put(6.0,0.6){\oval(1.0,1.0)\makebox(0,0){\bf \large {A}}}
\put(6.5,0.6){\vector(1,0){2}}
\put(9.5,0.6){\makebox(0,0){\bf \large {$\Phi$}}}
\end{picture}
\caption{The reactions taking place in a system of proliferating cells. 1) Whenever an $A$ agent meets an $S$ agent, a new $A$ agent is created. $A + S \rightarrow  S + A + A$, with a rate of $d_1$ 2) Each A agent has a constant probability $d_2$ of dying. $A \rightarrow \Phi $. The Ordinary Differential Equation (ODE) approximation in the continuum limit for these 2 reactions is: 
$\dot{A} = (<S>*d_1-d_2)*A$. In the left figure full arrow represent activation, and dotted arrows represent destruction.}
\end{figure}

\newpage

\begin{figure}
\includegraphics[clip,width=11 cm]{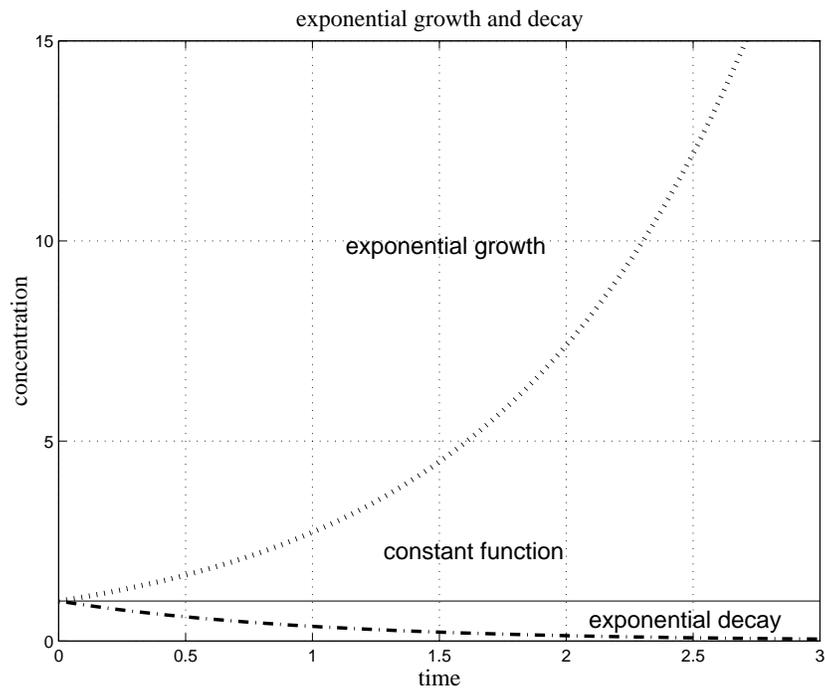}
\caption{The solutions of equation 3. An exponentially decaying solution for $(<S>*d_1-d_2) < 0 $ , an exponentially growing solution for $(<S>*d_1-d_2) > 0 $ and a constant solution for $(<S>*d_1-d_2) = 0 $.}
\end{figure}

\newpage

\begin{figure}
\includegraphics[clip,width=11 cm]{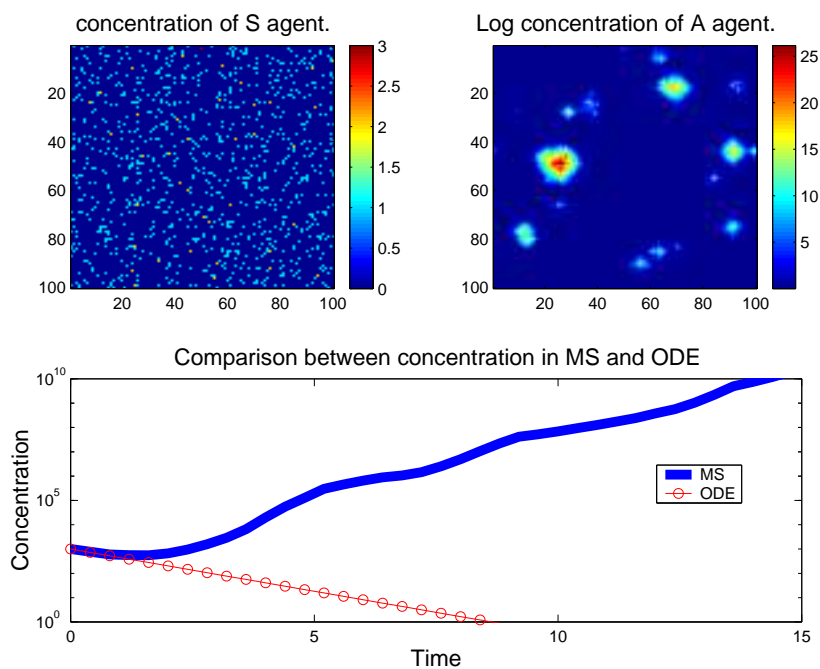}
\caption{Comparison between the MS and ODE description of the system described in figure 1. The A drawing is the local concentration of $S$ agents at every latice site. The B drawing represents the log of the $A$ agent concentration at every site. We use a log representation in order to show the large variance in the local concentration of the A agents. The C drawing shows the average $A$ agent concentration in the MS (interrupted lines)  and in the differential equation model (circles).}
\end{figure}

\newpage

\begin{figure}
\includegraphics[clip,width=11 cm]{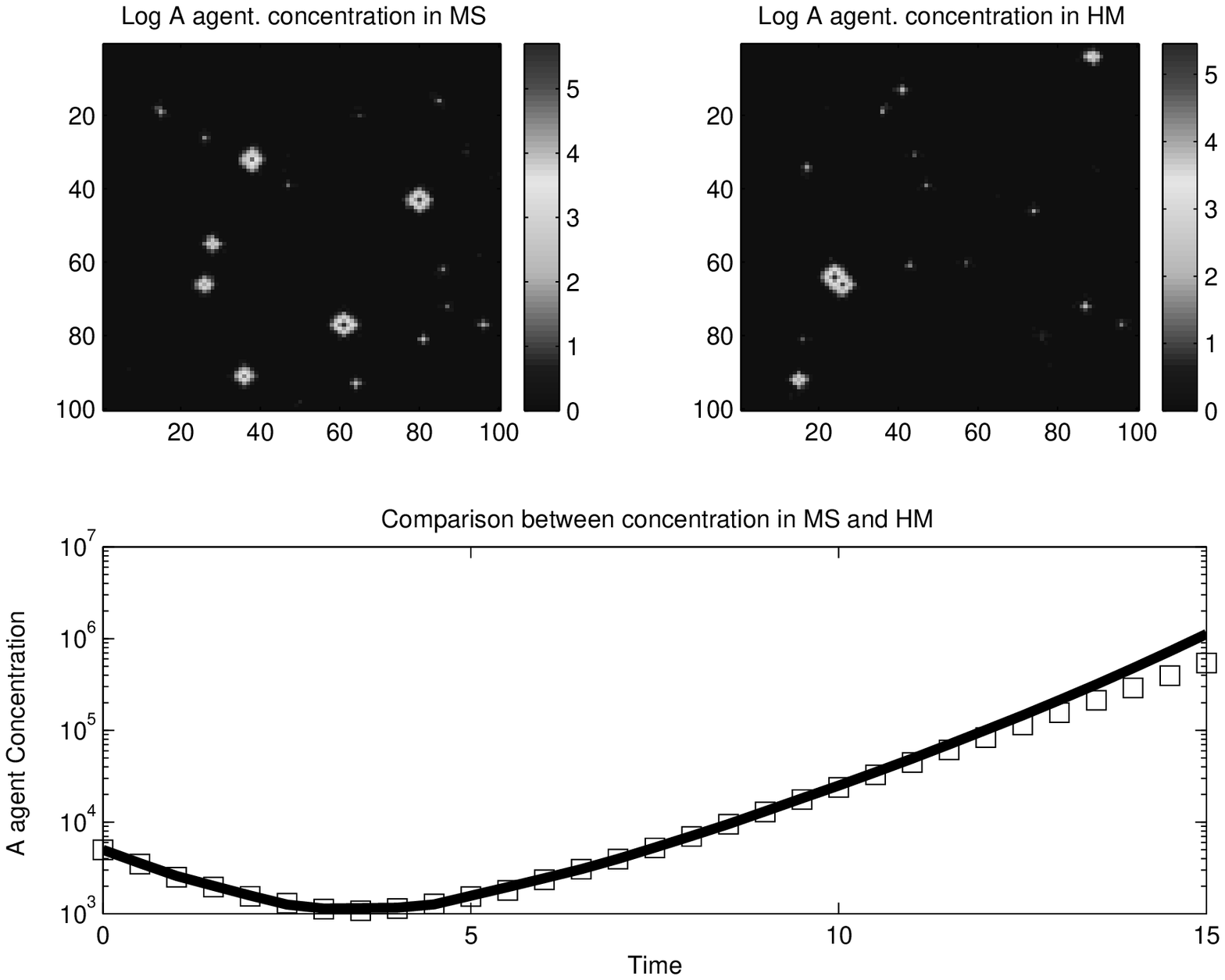}
\includegraphics[clip,width=5 cm]{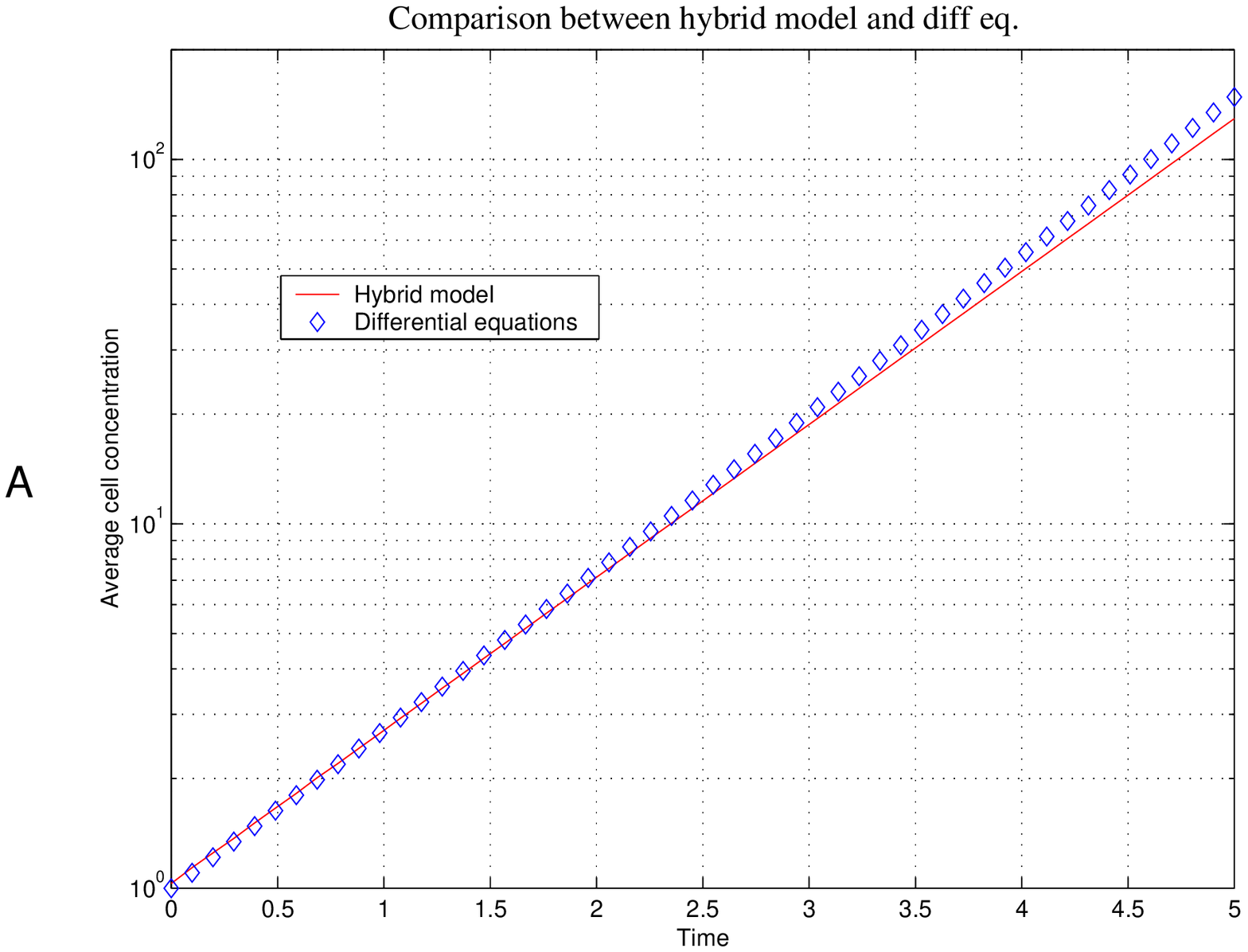}
\includegraphics[clip,width=5 cm]{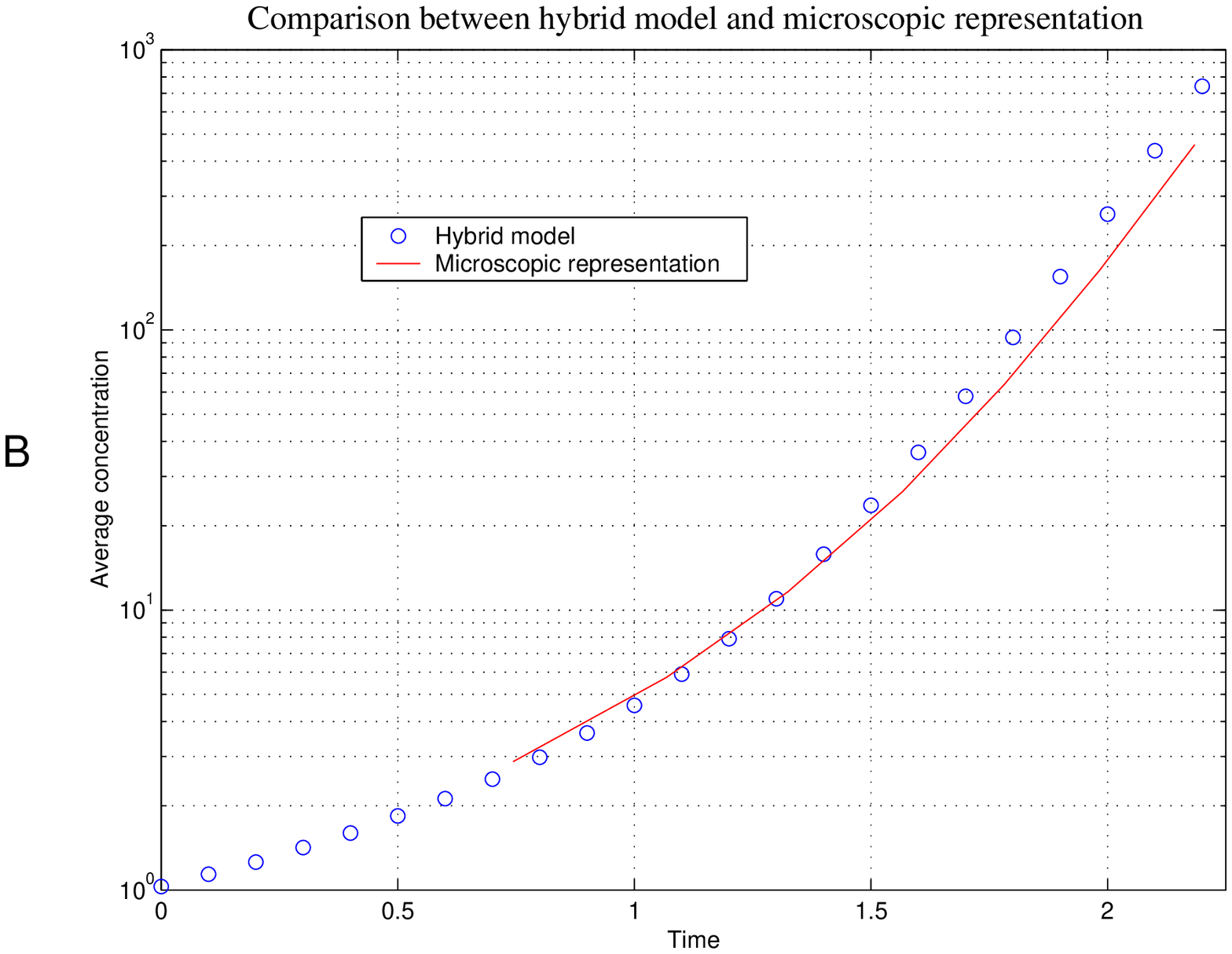}
\caption{A comparison between the concentration of $A$ agent in a MS and in  HM model. One can see that the concentration is similar in the 2 models, if one uses appropriate parameters.Comparison between a HM, an MS and an ODE. In the lower drawing we show that hybrid models can produce both the results of the microscopic representation, and the results of the differential equations. Figure A is the comparison between ODE and a HM for an homogeneous $S$ concentration. Figure 4B is the comparison between an MS and a HM for a non-homogeneous $S$concentration.}
\end{figure} 

\newpage

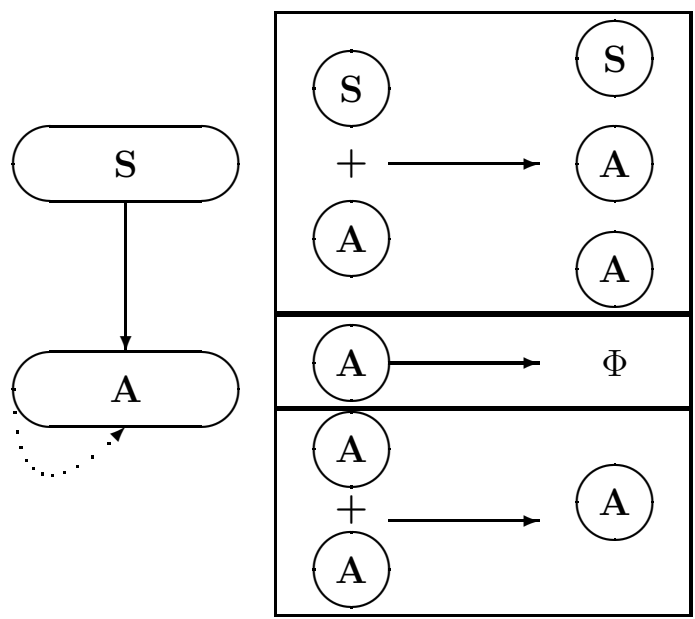
\begin{figure}
\setlength{\unitlength}{1cm}
\begin{picture}(11.0,8.0)
\thicklines
\put(3.0,6.0){\oval(3.0,1.0)\makebox(0,0){\bf \large {S}}}
\put(3.0,3.0){\oval(3.0,1.0)\makebox(0,0){\bf \large {A}}}
\put(3.0,5.5){\vector(0,-1){2}}
\bezier{12}(1.5,3.0)(1.5,1.0)(3.0,2.5)
\put(3.0,2.5){\vector(1,1){0}}

\put(5.0,4.0){\framebox(5.5,4.0){}}
\put(6.0,7.0){\oval(1.0,1.0)\makebox(0,0){\bf \large {S}}}
\put(6.0,5.0){\oval(1.0,1.0)\makebox(0,0){\bf \large {A}}}
\put(9.5,7.4){\oval(1.0,1.0)\makebox(0,0){\bf \large {S}}}
\put(9.5,6.0){\oval(1.0,1.0)\makebox(0,0){\bf \large {A}}}
\put(9.5,4.6){\oval(1.0,1.0)\makebox(0,0){\bf \large {A}}}
\put(6.0,6.0){\makebox(0,0){\bf \large {+}}}
\put(6.5,6.0){\vector(1,0){2}}
\put(5.0,2.75){\framebox(5.5,1.25){}}
\put(6.0,3.35){\oval(1.0,1.0)\makebox(0,0){\bf \large {A}}}
\put(6.5,3.35){\vector(1,0){2}}
\put(9.5,3.35){\makebox(0,0){\bf \large {$\Phi$}}}
\put(5.0,0.0){\framebox(5.5,2.75){}}
\put(6.0,2.2){\oval(1.0,1.0)\makebox(0,0){\bf \large {A}}}
\put(6.0,0.6){\oval(1.0,1.0)\makebox(0,0){\bf \large {A}}}
\put(6.0,1.4){\makebox(0,0){\bf \large {+}}}
\put(9.5,1.5){\oval(1.0,1.0)\makebox(0,0){\bf \large {A}}}
\put(6.5,1.25){\vector(1,0){2}}
\end{picture}
\caption{ Reaction scheme for a system containing local competition. This system has constant proliferation and death rates as in Figure 1. However it has an extra mechanism of local competition: Whenever two $A$ agents meet there is a probability that one of them will be destroyed $A + A -> A$}
\end{figure}

\newpage

\begin{figure}
\center
\noindent

\includegraphics[clip,width=11 cm]{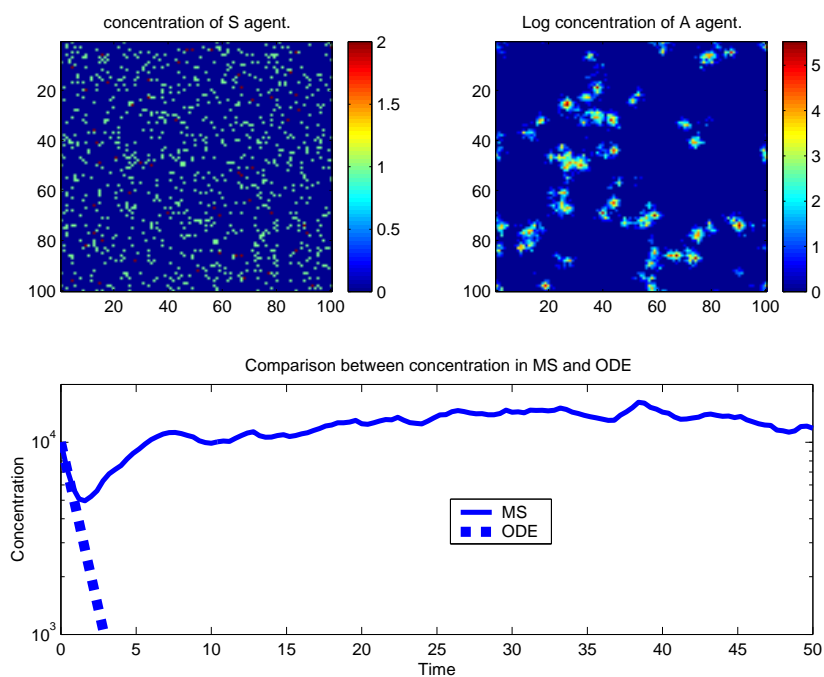}
\caption{Comparison between ODE and MS for the system defined by figure 5. This system has an average negative growth rate yielding an exponential decay in the ODE description as in the previous section. In the MS we observe a limited growth of the $A$ islands, which is due to the competition between $A$s. Thus even regions in which the local $A$ concentration would grow are limited to a finite concentration.}
\end{figure}

\newpage

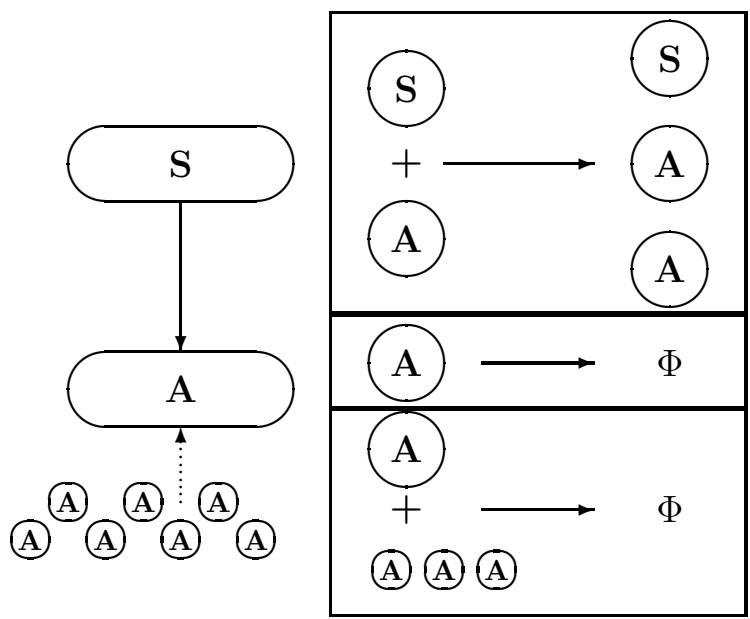
\begin{figure}
\setlength{\unitlength}{1cm}
\begin{picture}(11.0,8.0)
\thicklines
\put(3.0,6.0){\oval(3.0,1.0)\makebox(0,0){\bf \large {S}}}
\put(3.0,3.0){\oval(3.0,1.0)\makebox(0,0){\bf \large {A}}}
\put(3.0,5.5){\vector(0,-1){2}}
\put(3.5,1.5){\oval(0.5,0.5)\makebox(0,0){\bf \small {A}}}
\put(2.5,1.5){\oval(0.5,0.5)\makebox(0,0){\bf \small {A}}}
\put(1.5,1.5){\oval(0.5,0.5)\makebox(0,0){\bf \small {A}}}
\put(4.0,1.0){\oval(0.5,0.5)\makebox(0,0){\bf \small {A}}}
\put(3.0,1.0){\oval(0.5,0.5)\makebox(0,0){\bf \small {A}}}
\put(2.0,1.0){\oval(0.5,0.5)\makebox(0,0){\bf \small {A}}}
\put(1.0,1.0){\oval(0.5,0.5)\makebox(0,0){\bf \small {A}}}

\multiput(3.0,1.5)(0.0,0.1){10}{\circle*{0.05}}
\bezier{12}(2.0,2.5)(2.5,2.5)(3.0,2.5)
\put(3.0,2.5){\vector(0,1){0}}

\put(5.0,4.0){\framebox(5.5,4.0){}}
\put(6.0,7.0){\oval(1.0,1.0)\makebox(0,0){\bf \large {S}}}
\put(6.0,5.0){\oval(1.0,1.0)\makebox(0,0){\bf \large {A}}}
\put(9.5,7.4){\oval(1.0,1.0)\makebox(0,0){\bf \large {S}}}
\put(9.5,6.0){\oval(1.0,1.0)\makebox(0,0){\bf \large {A}}}
\put(9.5,4.6){\oval(1.0,1.0)\makebox(0,0){\bf \large {A}}}
\put(6.0,6.0){\makebox(0,0){\bf \large {+}}}
\put(6.5,6.0){\vector(1,0){2}}

\put(5.0,2.75){\framebox(5.5,1.25){}}
\put(6.0,3.35){\oval(1.0,1.0)\makebox(0,0){\bf \large {A}}}
\put(7.0,3.35){\vector(1,0){1.5}}
\put(9.5,3.35){\makebox(0,0){\bf \large {$\Phi$}}}

\put(5.0,0.0){\framebox(5.5,2.75){}}
\put(6.0,2.2){\oval(1.0,1.0)\makebox(0,0){\bf \large {A}}}
\put(6.0,1.4){\makebox(0,0){\bf \large {+}}}
\put(5.8,0.6){\oval(0.5,0.5)\makebox(0,0){\bf \small {A}}}
\put(6.5,0.6){\oval(0.5,0.5)\makebox(0,0){\bf \small {A}}}
\put(7.2,0.6){\oval(0.5,0.5)\makebox(0,0){\bf \small {A}}}
\put(9.5,1.4){\makebox(0,0){\bf \large {$\Phi$}}}
\put(7.0,1.4){\vector(1,0){1.5}}

\end{picture}
\caption{ Global competition reaction scheme. The mechanisms of proliferation and death are similar in this system to the one described in figure 1. The extra mechanism in this system is the global competition: Each $A$ agent has an extra death rate proportional to the average concentration of the $A$s.}
\end{figure}

\newpage

\begin{figure}
\includegraphics[clip,width=11 cm]{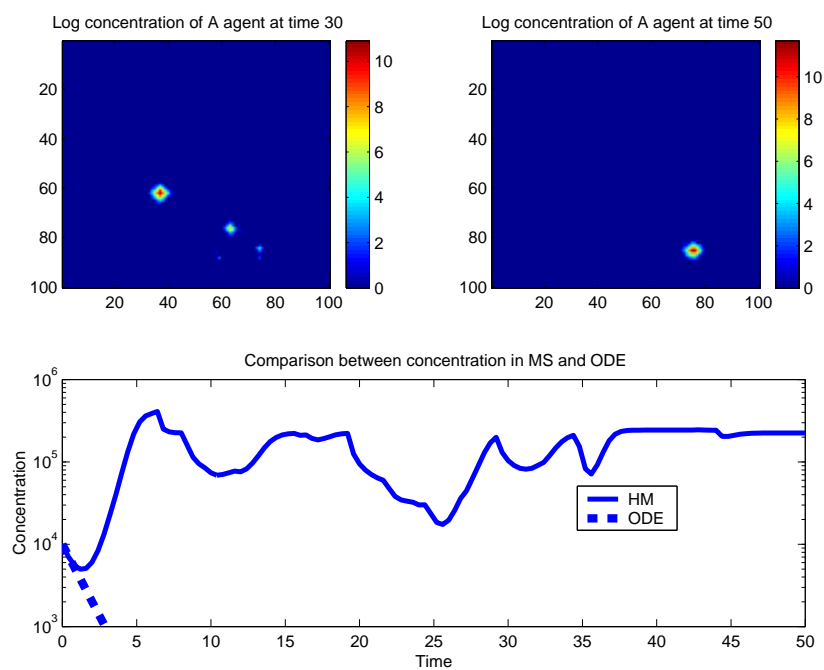}
\caption{ Comparison between ODE and HM for a system with an average negative growth rate and a global competition term (as defined in figure 7). In this model a single large island with a very high $A$ concentration appears. The competition with this  island destroys all the other small $A$ islands. The A and B drawings show the log of the $A$ agents concentration at different times. The C drawing shows the comparison between the ODE and the HM.}
\end{figure} 

\newpage

\begin{figure}
\includegraphics[clip,width=11 cm]{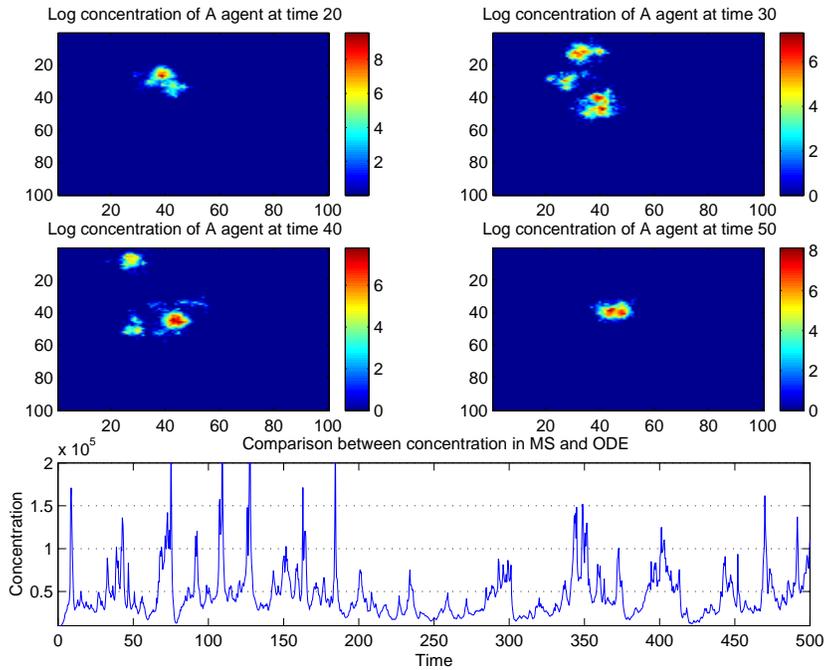}
\caption{The model defined in figure 7 in a regime with high $S$ diffusion rate. When the $S$ diffusion rate is very high, a few large $A$ islands are created. As opossed to figure 8, the largest $A$ island does not dominate. The various islands emerge an adaptive behavior: They look for the maxima of the $S$ concentration, split, merge and die. The top windows shows different snapshots of the $A$ distribution. The small islands are destroyed while the large islands grow. The bottom window is the a long term evolution of this system showing intermitent fluctuations.}
\end{figure} 

\newpage

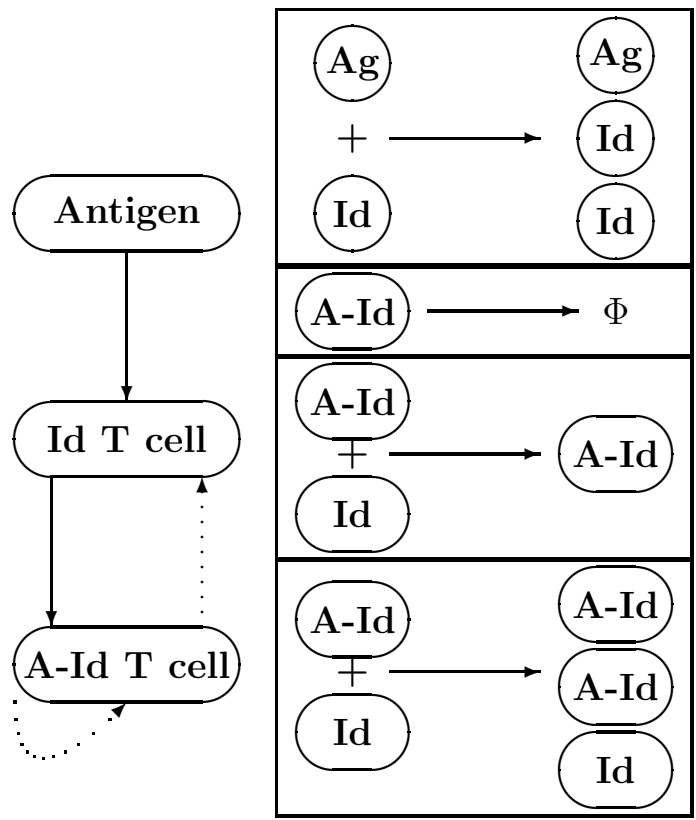
\begin{figure}
\setlength{\unitlength}{1cm}
\begin{picture}(11.0,11.0)
\thicklines
\put(3.0,8.0){\oval(3.0,1.0)\makebox(0,0){\bf \large {Antigen}}}
\put(3.0,5.0){\oval(3.0,1.0)\makebox(0,0){\bf \large {Id T cell}}}
\put(3.0,2.0){\oval(3.0,1.0)\makebox(0,0){\bf \large {A-Id T cell}}}
\multiput(4.0,2.5)(0.0,0.2){10}{\circle*{0.05}}
\put(4.0,4.5){\vector(0,1){0}}
\put(3.0,7.5){\vector(0,-1){2}}
\put(2.0,4.5){\vector(0,-1){2}}
\bezier{12}(1.5,1.5)(1.5,0.0)(3.0,1.5)
\put(3.0,1.5){\vector(1,1){0}}

\put(5.0,7.3){\framebox(5.5,3.4){}}
\put(6.0,10.0){\oval(1.0,1.0)\makebox(0,0){\bf \large {Ag}}}
\put(6.0,8.0){\oval(1.0,1.0)\makebox(0,0){\bf \large {Id}}}
\put(9.5,10.1){\oval(1.0,1.0)\makebox(0,0){\bf \large {Ag}}}
\put(9.5,9.0){\oval(1.0,1.0)\makebox(0,0){\bf \large {Id}}}
\put(9.5,7.9){\oval(1.0,1.0)\makebox(0,0){\bf \large {Id}}}
\put(6.0,9.0){\makebox(0,0){\bf \large {+}}}
\put(6.5,9.0){\vector(1,0){2}}

\put(5.0,6.1){\framebox(5.5,1.2){}}
\put(6.0,6.7){\oval(1.5,1.0)\makebox(0,0){\bf \large {A-Id}}}
\put(9.5,6.7){\makebox(0,0){\bf \large {$\Phi$}}}
\put(7.0,6.7){\vector(1,0){2}}

\put(5.0,3.4){\framebox(5.5,2.7){}}
\put(6.0,5.5){\oval(1.5,1.0)\makebox(0,0){\bf \large {A-Id}}}
\put(6.0,4.0){\oval(1.5,1.0)\makebox(0,0){\bf \large {Id}}}
\put(9.5,4.8){\oval(1.5,1.0)\makebox(0,0){\bf \large {A-Id}}}
\put(6.0,4.8){\makebox(0,0){\bf \large {+}}}
\put(6.5,4.8){\vector(1,0){2}}

\put(5.0,0.0){\framebox(5.5,3.4){}}
\put(6.0,2.6){\oval(1.5,1.0)\makebox(0,0){\bf \large {A-Id}}}
\put(6.0,1.1){\oval(1.5,1.0)\makebox(0,0){\bf \large {Id}}}
\put(9.5,2.8){\oval(1.5,1.0)\makebox(0,0){\bf \large {A-Id}}}
\put(9.5,1.7){\oval(1.5,1.0)\makebox(0,0){\bf \large {A-Id}}}
\put(9.5,0.6){\oval(1.5,1.0)\makebox(0,0){\bf \large {Id}}}
\put(6.0,1.9){\makebox(0,0){\bf \large {+}}}
\put(6.5,1.9){\vector(1,0){2}}
\end{picture}

\caption{The reactions scheme of a system describing the destruction of a pathogen by the immune system.}\end{figure}

\newpage

\begin{figure}
\includegraphics[clip,width=11 cm]{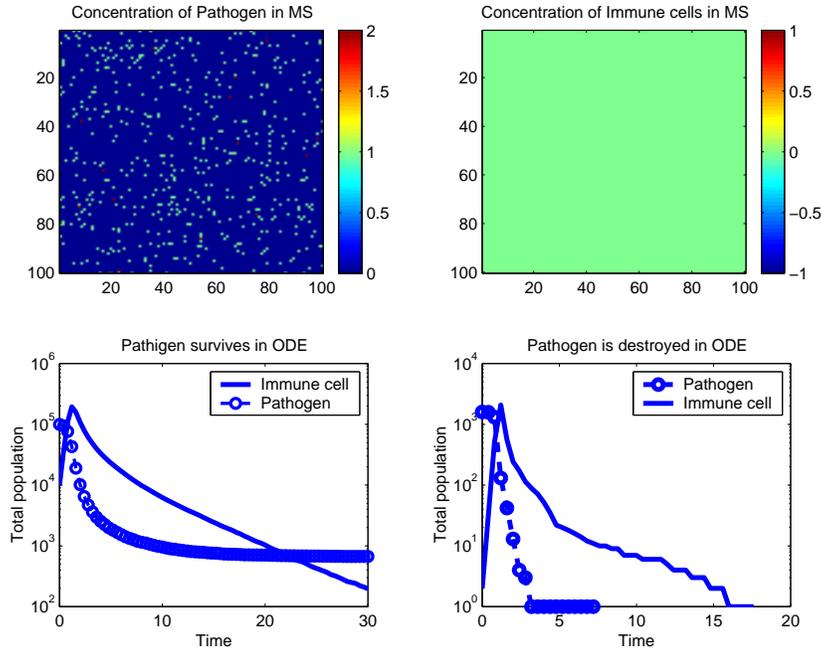}
\caption{ The simulation by a MS of a system describing the down regulation of pathogen by immune cells. In contradiction with the ODE description that predicts an uniform steady state, the MS predicts the formation of protected zones in which the pathogen can survive. The C and D drawings represents 2 cases predicted by the ODE description: In the C drawing the pathogen is not destroyed, and in the D drawing the immune system manages to destroy the pathogen. The A drawing describes the pathogen distribution simluated by the MS. One can see it contains protected regions containing pathogens surrounded by empty regions.}
\end{figure} 

\newpage

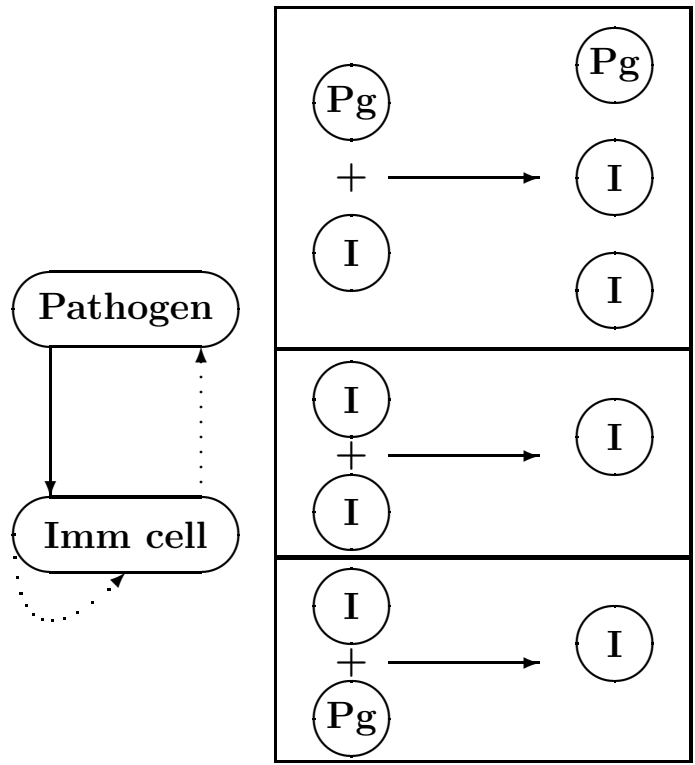
\begin{figure}
\setlength{\unitlength}{1cm}
\begin{picture}(6.0,10.0)
\thicklines
\put(3.0,6.0){\oval(3.0,1.0)\makebox(0,0){\bf \large {Pathogen}}}
\put(3.0,3.0){\oval(3.0,1.0)\makebox(0,0){\bf \large {Imm cell}}}
\put(2.0,5.5){\vector(0,-1){2}}
\multiput(4,3.5)(0.0,0.2){10}{\circle*{0.05}}
\put(4.0,5.5){\vector(0,1){0}}
\bezier{12}(1.5,3.0)(1.5,1.0)(3.0,2.5)
\put(3.0,2.5){\vector(1,1){0}}

\put(5.0,5.5){\framebox(5.5,4.5){}}
\put(6.0,8.75){\oval(1.0,1.0)\makebox(0,0){\bf \large {Pg}}}
\put(6.0,6.75){\oval(1.0,1.0)\makebox(0,0){\bf \large {I}}}
\put(9.5,9.25){\oval(1.0,1.0)\makebox(0,0){\bf \large {Pg}}}
\put(9.5,7.75){\oval(1.0,1.0)\makebox(0,0){\bf \large {I}}}
\put(9.5,6.25){\oval(1.0,1.0)\makebox(0,0){\bf \large {I}}}
\put(6.0,7.75){\makebox(0,0){\bf \large {+}}}
\put(6.5,7.75){\vector(1,0){2}}

\put(5.0,2.7){\framebox(5.5,2.75){}}
\put(6.0,4.8){\oval(1.0,1.0)\makebox(0,0){\bf \large {I}}}
\put(6.0,3.3){\oval(1.0,1.0)\makebox(0,0){\bf \large {I}}}
\put(9.5,4.3){\oval(1.0,1.0)\makebox(0,0){\bf \large {I}}}
\put(6.0,4.05){\makebox(0,0){\bf \large {+}}}
\put(6.5,4.05){\vector(1,0){2}}

\put(5.0,0.0){\framebox(5.5,2.7){}}
\put(6.0,2.05){\oval(1.0,1.0)\makebox(0,0){\bf \large {I}}}
\put(6.0,0.55){\oval(1.0,1.0)\makebox(0,0){\bf \large {Pg}}}
\put(9.5,1.55){\oval(1.0,1.0)\makebox(0,0){\bf \large {I}}}
\put(6.0,1.3){\makebox(0,0){\bf \large {+}}}
\put(6.5,1.3){\vector(1,0){2}}

\end{picture}

\caption{ Reaction scheme for a system describing the regulation of idiotypic cells by anti-idiotypic cells.}
\end{figure}

\newpage

\begin{figure}
\includegraphics[clip,width=11 cm]{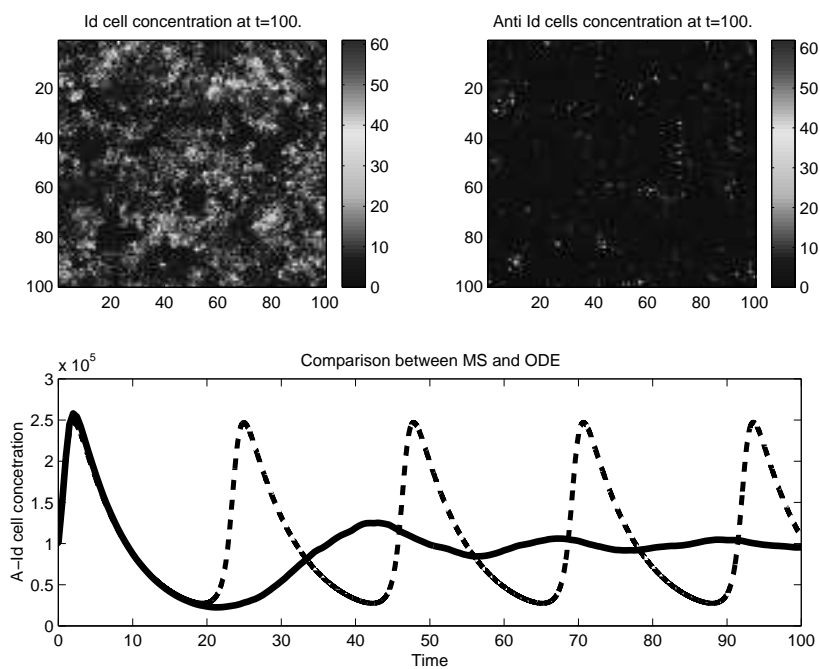}
\caption{A comparison between the MS and the ODE for a system containing idiotypic and anti idiotypic cells. The ODE leads to oscilations. These oscialtions disapear in the MS when the random reactions are taken into acount. The MS predicts that the Id and A-Id cells concentration decays to a constant.}
\end{figure} 

\newpage

\begin{figure}
\includegraphics[clip,width=11 cm]{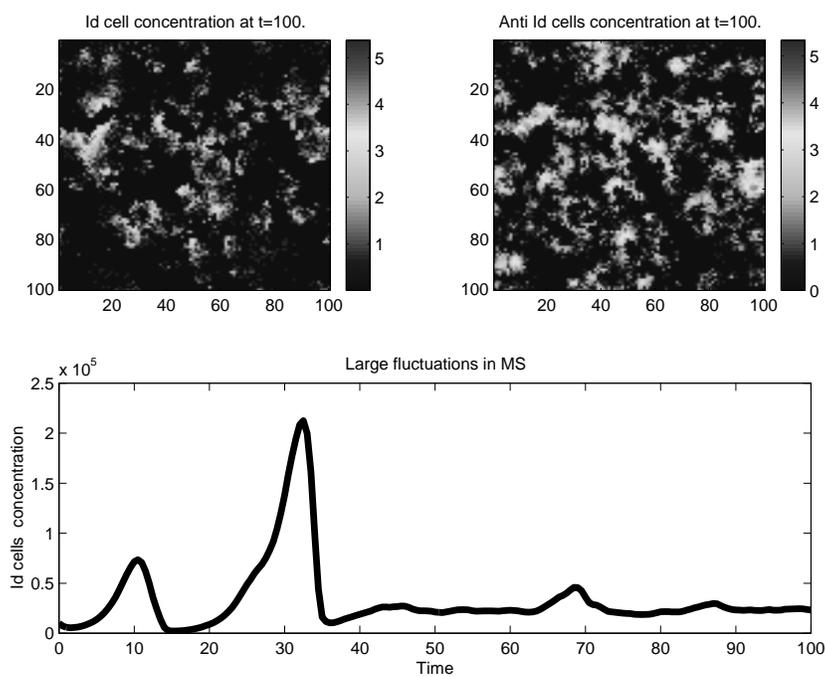}
\caption{Large fluctuations can appear in the MS description of a system with Id and anti Id cells. Most of the contribution for the concentration is in very localized islands of very high concentration. These leads to the very sharp fluctuations. This system reaches a final steady state in which space is occupied by large disjoint regions of Id and respectively  A-Id cells. Each such region has non stationary dynamics.}
\end{figure} 

\end{document}